\documentclass[12pt]{article}
\usepackage{amsfonts}
\usepackage{amsmath}
\usepackage{amssymb}
\usepackage{graphicx}
\usepackage{color}
\usepackage[all, knot]{xy}
\usepackage{tikz}

\usepackage{ulem}

\usepackage[utf8]{inputenc}
\usepackage{epstopdf}
\usepackage[footnotesize]{caption}
\usepackage{amsthm}
\usepackage{enumitem}
\usepackage{mathrsfs}

\usepackage[margin=3cm]{geometry}

\def \be {\begin{equation}}
\def \ee {\end{equation}}
\def \bea {\begin{eqnarray}}
\def \eea {\end{eqnarray}}
\def \nn {\nonumber}

\def \rr {\raise.35ex\hbox{\small $\prime$}\kern-.17em{\mbox{\large $\imath$}}}

\def \dels {\partial\kern-.6em /\kern.1em}
\def \As {{A\kern-.5em / \kern.5em}}
\def \Ds {D\kern-.7em / \kern.5em}

\def \ks {k\kern-.5em /}
\def \ls {l\kern-.5em /}

\newcommand{\er}[1]{\eqref{#1}}

\newcommand{\ci}[1]{}

\newcommand{\lb}{\left(}
\newcommand{\rb}{\right)}

\newcommand{\p}{\partial}

\newcommand{\ba}{\begin{eqnarray}}
\newcommand{\ea}{\end{eqnarray}}
\newcommand{\bal}{\begin{align}}
\newcommand{\eal}{\end{align}}
\newcommand{\bay}[1]{\left(\begin{array}{#1}}
\newcommand{\eay}{\end{array}\right)}

\newcommand{\at}[1]{{\Big|}_{#1}}

\def\xd{{\delta}}

\def\xl{{\lambda}}

\newcommand{\hide}[1]{}

\newlist{axioms}{enumerate}{2}
\setlist[axioms,1]{label=\textbf{A\arabic{axiomsi}.}, ref=A\arabic{axiomsi}}
\setlist[axioms,2]{label=\textbf{A\arabic{axiomsi}\rlap{\myEnumCounter{axiomsii}}.},%
                   ref=A\arabic{axiomsi}\myEnumCounter{axiomsii},%
                   align=parleft,%
                   leftmargin=0em,%
                   itemsep=1.4ex,%
                   before={\stepcounter{axiomsi}}}

  \usetikzlibrary{decorations.markings}

\begin{document}

\begin{titlepage}
\begin{center}

\textbf{\LARGE
Berry Curvature and\\
 Riemann Curvature in Kinematic Space\\ 
 with Spherical Entangling Surface
\vskip.3cm
}
\vskip .5in
{\large
Xing Huang$^{a,b,c}$ \footnote{e-mail address: xingavatar@gmail.com} and
Chen-Te Ma$^{d,e,f,g}$ \footnote{e-mail address: yefgst@gmail.com}
\\
\vskip 1mm
}
{\sl
$^a$
Institute of Modern Physics, Northwest University, Xian 710069, China.
\\
$^b$
Shaanxi Key Laboratory for Theoretical Physics Frontiers, Xian 710069, China.
\\
$^c$
NSFC-SPTP Peng Huanwu Center for Fundamental Theory, Xian 710127, China.
\\
$^d$
Guangdong Provincial Key Laboratory of Nuclear Science,\\
 Institute of Quantum Matter,
South China Normal University, Guangzhou 510006, Guangdong, China.
\\
$^e$
School of Physics and Telecommunication Engineering,\\
South China Normal University, Guangzhou 510006, Guangdong, China.
\\
$^f$
The Laboratory for Quantum Gravity and Strings,\\
 Department of Mathematics and Applied Mathematics,\\
University of Cape Town, Private Bag, Rondebosch 7700, South Africa.
\\
$^g$
Department of Physics and Center for Theoretical Sciences, \\
National Taiwan University, Taipei 10617, Taiwan, R.O.C..
}\\
\vskip 1mm
\vspace{40pt}
\end{center}
\newpage
\begin{abstract}
We discover the connection between the Berry curvature and the Riemann curvature tensor in any kinematic space of minimal surfaces anchored on spherical entangling surfaces. This new holographic principle establishes the Riemann geometry in kinematic space of arbitrary dimensions from the holonomy of modular Hamiltonian, which in the higher dimensions is specified by a pair of time-like separated points as in CFT$_1$ and CFT$_2$. 
The Berry curvature that we constructed also shares the same property of the Riemann curvature for all geometry: internal symmetry; skew symmetry; first Bianchi identity.
We derive the algebra of the modular Hamiltonian and its deformation, the
latter of which can provide the maximal modular chaos to the modular scrambling modes.
The algebra also dictates the parallel transport, which leads
to the Berry curvature exactly matching to the Riemann curvature tensor. Finally, we compare CFT$_1$ to higher dimensional CFTs and show the difference from the OPE block.
\end{abstract}
\end{titlepage}

\section{Introduction}
\label{sec:1}
\noindent
{\it Holographic principle} states that the physical degrees of freedom in quantum gravity \cite{tHooft:1996rdg} is encoded on its boundary. 
One difficulty in studying quantum gravity is the non-renormalizability problem in Einstein gravity theory \cite{Strominger:2017zoo}. 
The principle avoids the problem by a boundary theory with a flat background. 
String theory is the only known candidate for perturbative quantum gravity,
and it leads to the {\it holographic principle} of {\it Anti-de Sitter/Conformal Field Theory} (AdS/CFT) correspondence \cite{Maldacena:1997re}.  
The correspondence is still a conjecture, but it is convincing from practical calculations for various cases. 
The success of the AdS/CFT correspondence suggests a probe of quantum gravity from {\it emergent spacetime}.
\\

\noindent
The philosophy of the emergent spacetime is to obtain an equivalent description of spacetime from other objects. 
Based on the holographic principle, boundary theory is naturally expected to {\it reconstruct} bulk gravity theory. 
To understand the details, knowing what geometric objects {\it emerges} from becomes the first important task. 
One interesting observation of the emergent geometry is entanglement entropy in CFT$_d$ between the subregion and the complement region can be given by a co-dimensional two surface with a minimal area in AdS$_{d+1}$ (minimum surface) \cite{Ryu:2006bv}. 
This observation implies that the mechanism of generating spacetime should be {\it quantum mechanical} because the {\it geometric} object is related to {\it quantum information} quantity. 
Hence studying {\it quantum information} in a non-gravity theory should be the same as studying gravity and is even simpler. 
\\

\noindent
Using a conformal transformation, entanglement entropy with a spherical entangling surface can be translated to thermal entropy \cite{Casini:2011kv}. 
The AdS/CFT correspondence gives a consistent result to the conjecture of holographic entanglement entropy \cite{Casini:2011kv}. 
Calculation of entanglement entropy in quantum field theory is
usually based on a replica trick, which is generic but hard \cite{Ma:2015xes, Huang:2016bkp, Ma:2016deg, Ma:2016pah}. 
The minimum surface gives a simple way of obtaining an exact formula for entanglement entropy. 
Developing practical methods for quantum information quantities in a strongly coupled system is meaningful \cite{Ma:2018efs}. 
\\

\noindent
A linearized perturbation of the minimum surface is dual to the {\it OPE block} of a stress tensor in pure AdS \cite{Czech:2016xec, deBoer:2016pqk}. 
The OPE block of a stress tensor also corresponds to a {\it modular Hamiltonian} \cite{Casini:2011kv}. Hence the {\it modular Hamiltonian} \cite{Witten:2018lha} is connected to the minimum surface. 
A probe of the bulk gravity theory is necessary to build an operator dictionary. 
To reach the goal, it is convenient to relate quantum information quantities to correlation functions \cite{Dolan:2003hv, SimmonsDuffin:2012uy, Fitzpatrick:2015foa, daCunha:2016crm, Guica:2016pid, Fitzpatrick:2016mtp}. 
The OPE block provides such a connection. 
The OPE block is to use the operator product expansion to organize bi-local operators \cite{Hijano:2015zsa}. 
The geometry of the {\it kinematic space} was obtained from a scalar OPE block, which follows the Klein-Gordon equation \cite{Czech:2016xec, deBoer:2016pqk}. 
In this paper, the kinematic space is only for co-dimension two minimal surfaces
associated with spherical entangling surfaces.
\\

\noindent
For CFT$_d$, the dimensions of the kinematic space are $2d$. 
The geometry of the kinematic space
can be determined by symmetry but physical meanings of various quantities
like {\it connection}, {\it Riemann curvature tensor}, etc, are less known and worth a further study. 
Hence constructing the geometric objects in the kinematic space from quantum field theory is non-trivial.
Recently, using the algebra of a modular Hamiltonian in CFT$_2$, people derived the {\it modular Berry connection}, which leads to the {\it Berry curvature} \cite{Czech:2017zfq, Czech:2019vih}. 
In Lorentzian CFT$_1$, it was shown that the {\it Berry curvature} is equivalent to the {\it Riemann curvature tensor} \cite{Huang:2019wzc}. 
This study provides the {\it Riemann geometry} to the kinematic space from the perspective of quantum field theory.
\\

\noindent 
In this paper, we only consider a Lorentzian manifold.
The kinematic space in Lorentzian CFT$_1$ was not discussed often because the bulk object is not related to any reduced density matrix associated
to a spatial subregion \cite{Maldacena:2016hyu, Callebaut:2018xfu, Blommaert:2019hjr}.
Recently, the authors proposed that using the {\it OPE block} of a {\it stress tensor} to define the {\it modular Hamiltonian} should be useful for a study of AdS/CFT correspondence although it possibly loses the connection with entanglement \cite{Huang:2019wzc}.
\\

\noindent
The Berry connection gives a direct reconstruction of bulk spacetime from a quantum information perspective. 
Quantum entanglement provides information about how to entangle two subregions. 
Quantum entanglement should not be enough for the reconstruction of spacetime because it only refers to relate to other subsystems without any dynamic information about itself. 
The goal of quantum chaos is to distinguish integrable and chaotic quantum systems. 
Studying such a subject from a holographic perspective should provide additional information to the {\it emergent spacetime}. 
Recently, one quantum chaotic phenomenon, {\it sensitivity on the initial condition}, was studied holographically \cite{Shenker:2013pqa}. 
Direct computation in boundary theories also showed such a quantum chaotic phenomenon \cite{Maldacena:2015waa, Perlmutter:2016pkf, Jensen:2016pah}. 
These studies motivate a conjecture for a holographic study of Einstein gravity theory from {\it maximal quantum chaos} \cite{Maldacena:2016upp}. 
The holographic study, however, seems to show no connection between quantum chaos and quantum entanglement.
More recently, the algebra of a modular Hamiltonian shows {\it its sensitivity on an initial deformation of a modular Hamiltonian} \cite{deBoer:2019uem}, which should provide information describing the dynamics of a system. 
Hence the {\it kinematic space} connects two important directions of quantum information, {\it quantum entanglement} and {\it quantum chaos}.   
\\

\begin{figure}
\begin{center}
\includegraphics[width=8cm]{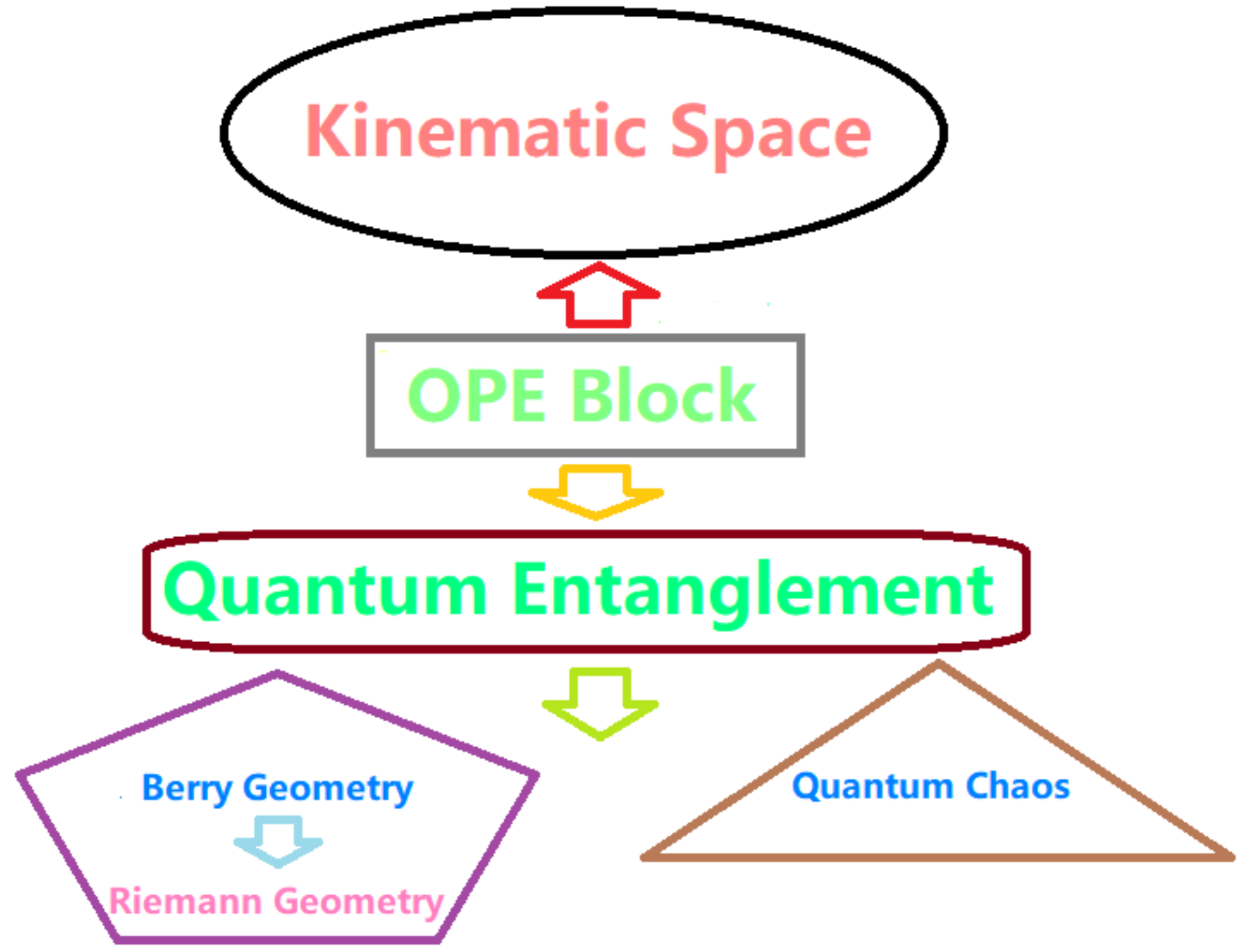}
\end{center}
\caption{The relation of our generalization.}
\label{Relation.pdf}
\end{figure}

\noindent
In this paper, we generalize the holographic principle of the kinematic space, associated with a spherical entangling surface in higher dimensional CFTs. 
The generalization gives the geometric objects of a modular Hamiltonian to supply the geometric objects of the kinematic space. The algebra of a modular Hamiltonian also connects quantum chaos to quantum entanglement in this generalization. 
The relations of above generalizations can be seen in Fig. \ref{Relation.pdf}.
To summarize our results:
\begin{itemize} 
\item We derive the algebra of a modular Hamiltonian with a spherical entangling surface and its deformation, which leads to the geometric pictures and maximal modular chaos as the modular scrambling modes for all CFTs. 
This gives a systematic derivation of the algebra for both 1d CFT \cite{Huang:2019wzc} and higher-dimensional CFTs \cite{Casini:2017roe} for the first time. The connection of quantum entanglement to quantum chaos shows the usefulness of the definition of a modular Hamiltonian in CFT$_1$.
\item We construct the modular Berry connection and Berry curvature. The Berry curvature shows the internal symmetry, skew symmetry, and satisfies the
first Bianchi identity. This construction requires the geometry of a kinematic space to have a similar form of Riemann geometry. The modular Berry connection is used for the parallel transport, and, we
find that the Berry curvature is equivalent to the Riemann curvature tensor in the kinematic space. 
\item We obtain a solution for the OPE block of a stress tensor. 
The reconstruction of kinematic space in CFT$_1$ should show the special
value
of the OPE block of a stress tensor, defined as a modular Hamiltonian in this paper. 
Although the procedure of the reconstruction is similar in all dimensions, the operator correspondence should be different between CFT$_1$ and the higher dimensional CFTs as OPE block of a stress tensor in CFT$_1$ has no connection to a reduced density matrix. 
\end{itemize}

\noindent
The rest of the paper is organized as follows: We derive the algebra of a modular Hamiltonian with a spherical entangling surface and use the algebra to show the maximal modular scrambling and solve the equation of parallel transport in Sec.~\ref{sec:2}. 
The holonomy around an infinitesimal parallelogram can be expressed in terms of commutator of tangent vectors, which is shown in Sec.~\ref{sec:3}. 
We use the result of the commutator to construct the Berry curvature and show the equivalence between the Berry curvature and Riemann curvature tensor in Sec.~\ref{sec:4}. Discussion of the difference in CFT$_1$ and the higher-dimensional CFTs is shown in Sec.~\ref{sec:5}. 
In the end, we discuss and conclude in Sec.~\ref{sec:6}. 
We give the details of the derivation of the algebra of a modular Hamiltonian in Appendix \ref{appa}. 
The solution of an OPE block in CFT$_1$ is checked in Appendix \ref{appb}.

\section{Algebra of Modular Hamiltonian}
\label{sec:2}
\noindent
We consider the modular Hamiltonian of a spherical region $A$ specified by a pair of time-like separated points, $x^{\mu}$ and $y^{\mu}$ \cite{Casini:2011kv}. The algebra of the modular Hamiltonian and its deformation for these points implies that scrambling modes of the deformation lead to the maximum modular chaos. We also use the algebra to construct the generator of parallel transport. 
We provide details to the derivation of the algebra of the modular Hamiltonian in Appendix \ref{appa}.

\subsection{Modular Hamiltonian}
\noindent
The modular Hamiltonian is defined by
\bea
H_{\mathrm{mod}}\equiv-\ln\rho_A,
\eea
where $\rho_A$ is a reduced density matrix of a region $A$, when the dimensions of spacetime in boundary field theory are larger than one.
\\ 

\noindent
The modular Hamiltonian of a ($d-1$)-dimensional ball-shaped region $A$ in CFT$_d$ is \cite{Casini:2011kv}
\bea
H_{\mathrm{mod}}=\int_A d\Sigma^{\mu}\ T_{\mu\nu}K^{\nu},
\eea
in which the integration of the region $A$ runs over 
\bea
|\vec{x}-\vec{x}_0|^2\le R^2
\eea
on a fixed time slice, and $R$ is the radius of the sphere. The dimensions of the spacetime in CFT$_d$ is labeled by $\mu=0, 1, \cdots, d-1$. The $T_{\mu\nu}$ is a traceless stress tensor. We choose the conformal Killing vector $K^{\mu}$ as the below \cite{Casini:2011kv}
\bea
K^{\mu}(w)\partial_{\mu, w}
=-\frac{2\pi}{(y-x)^2}\big((y-w)^2(x^{\mu}-w^{\mu})-(x-w)^2(y^{\mu}-w^{\mu})\big)\partial_{\mu, w},
\eea
where 
\bea
(y-x)^2\equiv \eta_{\mu\nu}(y-x)^{\mu}(y-x)^{\nu}, \qquad \eta_{\mu\nu}\equiv\mathrm{diag}(-1, 1, 1, \cdots, 1);
\eea
\bea
\partial_{\mu, w}\equiv\frac{\partial}{\partial w^{\mu}}.
\eea
When we have:
\bea
w^{\mu}=x^{\mu};\ w^{\mu}=y^{\mu};\ (y-w)^2\ \mathrm{and}\ (x-w)^2=0,
\eea
the conformal Killing vector vanishes. Hence it preserves the causal diamond.

\subsection{Algebra}
\noindent
We can generate an algebra of the modular Hamiltonian from the following identification with the conformal Killing vector 
\bea
H_{\mathrm{mod}}\rightarrow i K^{\rho}\partial_{\rho, w}.
\eea 
The algebra of the modular Hamiltonian is:
 \bea
 \lbrack H_{\mathrm{mod}}, H_{\mathrm{mod}}\rbrack&=&0; 
 \nn\\
 \lbrack H_{\mathrm{mod}}, \partial_{\nu, x} H_{\mathrm{mod}}\rbrack&=&-2\pi i\partial_{\nu, x}H_{\mathrm{mod}};
 \nn\\
 \lbrack H_{\mathrm{mod}}, \partial_{\nu, y} H_{\mathrm{mod}}\rbrack&=&2\pi i\partial_{\nu, y}H_{\mathrm{mod}}. \label{algebradeform}
 \eea 
 The details of the deriving the algebra can be seen in Appendix. \ref{appa}.
 Note that the deformation of $H_{\mathrm{mod}}$ for tips $(x^\mu, y^\mu)$ of the causal diamond essentially becomes the null deformation, and the algebra above follows from those obtained in Ref. \cite{Casini:2017roe}, but the derivation only can be applied to $d>1$ because the entangling surface is 0d for CFT$_1$. 
 Recently, using the OPE block of a stress tensor ones can define the modular Hamiltonian in CFT$_1$ with a holographic correspondence \cite{Huang:2019wzc}. 
Hence we want to use the conformal Killing vector to provide a systematic derivation. Moreover, the details of Appendix \ref{appa} are also useful for obtaining $\lbrack\partial_{\mu, y}H_{\mathrm{mod}}, \partial_{\nu, x} H_{\mathrm{mod}}\rbrack$ that we will need.  
\\

\subsection{Maximal Modular Scrambling}
\noindent
 We can infinitesimally perturb the modular Hamiltonian, which can be done by deforming a region's shape or perturbing a quantum state,
 \bea
 &&\exp(-i H_{\mathrm{mod}}s)\exp\big(i(H_{\mathrm{mod}}+\epsilon\delta H_{\mathrm{mod}})s\big)
 \nn\\
&=&
 \exp\bigg(i\epsilon\int_0^sds^{\prime}\ \exp(-iH_{\mathrm{mod}} s^{\prime})\delta H_{\mathrm{mod}}\exp(i H_{\mathrm{mod}}s^{\prime})
 +{\cal O}(\epsilon^2)\bigg).
 \eea
 We then can find \cite{deBoer:2019uem}
 \bea
 \exp(-i H_{\mathrm{mod}}s)\delta H_{\mathrm{mod}}\exp(i H_{\mathrm{mod}}s)\sim \exp(2\pi s) \delta H_{\mathrm{mod}}
 \eea
 from 
 \bea
  \lbrack H_{\mathrm{mod}}, \partial_{y, \nu} H_{\mathrm{mod}}\rbrack=2\pi i\partial_{y, \nu}H_{\mathrm{mod}}.
  \label{cmdm}
 \eea
 Hence we find the exponent 
 \bea
 \lambda=2\pi,
 \eea
 which saturates the bound \cite{deBoer:2019uem}.
 \\
 
 \noindent
 Because we only use the algebra of the modular Hamiltonian, the saturation does not depend on any detail of CFT. Hence the maximal modular scrambling is not enough to distinguish chaotic theory from non-chaotic theory. Granted, it is hard to expect only kinematic information like algebra can provide useful constraint to a holographic study of Einstein gravity theory. Here we consider the deformation of the tips of a causal diamond. 
 Indeed, this choice of deformation is quite special. 
 Hence our study suggests that the additional information of emergent spacetime should come from other deformations. 
 \\
 
  \noindent
  Here the modular Hamiltonian in CFT$_1$ is defined by the conformal Killing vector. 
  Later we will introduce the OPE block \cite{Czech:2016xec, deBoer:2016pqk} of a stress tensor relating to this modular Hamiltonian.

 \subsection{Parallel Transport} 
\noindent
The parallel transport of the modular Hamiltonian is given by:
 \bea
 \partial_{\lambda}H_{\mathrm{mod}}=\lbrack V_{\delta\lambda}, H_{\mathrm{mod}}\rbrack; \qquad 
 P_0[V_{\delta\lambda}]=0,
 \label{eqparalleltr}
 \eea
 where
 \bea
 \partial_{\lambda}H_{\mathrm{mod}}
 &=&
 (\partial_{\lambda} x^{\mu})(\partial_{\mu, x} H_{\mathrm{mod}})
 +(\partial_{\lambda} y^{\mu})(\partial_{\mu, y} H_{\mathrm{mod}});
 \nn\\
 V_{\delta\lambda}&\equiv&\frac{1}{2\pi i} 
 \big((\partial_{\lambda} x^{\mu})(\partial_{\mu, x} H_{\mathrm{mod}})
 -(\partial_{\lambda} y^{\mu})(\partial_{\mu, y} H_{\mathrm{mod}})\big).
 \eea
 We define the projection operator to the space of zero modes of $H_{\mathrm{mod}}$ as $P_0 $. 
 The algebra \eqref{algebradeform} solves the equation \er{eqparalleltr} and gives the tangent vector $V_{\delta\lambda}$ of the lifted curve in the total space, which generates the parallel transport. 
 Note that as an eigenvector with non-vanishing eigenvalues, $V_{\delta \xl}$ lies in a different subspace compared to the zero modes (all with eigenvalues $0$), which is guaranteed by Eq. \eqref{algebradeform}.
 \\

\section{Berry Curvature from Holonomy}
\label{sec:3}
 \noindent
The holonomy measures the failure of closeness of the horizontal lift of a loop. The Berry curvature follows from the holonomy around an infinitesimal parallelogram. The solution $V_{\delta\lambda}$ to Eq. \er{eqparalleltr} gives the generator of parallel transport along a curve parameterized by $\xl$. It can be shown that the Berry curvature is given by $-\lbrack V_{\xd \xl_1},V_{\xd \xl_2}\rbrack$ \footnote{More precisely, for a parallelogram specified by tangent vectors $\xd \xl_1\p_{\xl_1},\xd \xl_2\p_{\xl_2}$ the holonomy around a lifted loop or simply the change of coordinate $X^\mu$ in the total space is given by 
\bea & \xd \xl_1 X_1 (\frac 1 2 \xd \xl_1X_1^\mu)+\xd \xl_2 X_2 (\xd \xl_1X_1^\mu+ \frac 1 2\xd \xl_2 X_2^\mu)-\xd \xl_1 X_1(\frac 1 2 \xd \xl_1X_1^\mu+\xd \xl_2 X_2^\mu)\nn\\ & -\xd \xl_2 X_2(\frac 1 2 \xd \xl_2 X_2^\mu) =  \xd \xl_1 \xd \xl_2 \lb (X_1)^\mu \p_{X^\mu} X_2 -(X_2)^\mu \p_{X^\mu} X_1\rb = \xd \xl_1 \xd \xl_2 [X_1, X_2]\,,\nn\eea 
where $X_1, X_2$ are the lifted vectors of $\p_{\xl_1}, \p_{\xl_2}$. On the other hand, we know that the change around such an infinitesimal parallelogram is related to the (Berry) curvature by $- \xd \xl_1 \xd \xl_2 {\cal R}(\p_{\xl_1},\p_{\xl_2})$}. We will subsequently calculate the commutator $-\lbrack V_{\mu, x}, V_{\nu, x}\rbrack$, $-\lbrack V_{\mu, y}, V_{\nu, y}\rbrack$, and $-\lbrack V_{\mu, x}, V_{\nu, y}\rbrack$ from $\lambda_{1,2}=x^{\mu}$ and $\lambda_{1,2}=y^{\mu}$ respectively.
\\
 
 \noindent
First, we have $-\lbrack V_{\mu, x}, V_{\nu, x}\rbrack$:
\bea
-\lbrack V_{\mu, x}, V_{\nu, x}\rbrack
 &\equiv& -\lbrack V_{\xd \xl_1}\at{\xl_1 = x^\mu}, V_{\xd \xl_2}\at{\xl_2 = x^\nu} \rbrack \nn\\
&=&
\frac{1}{4\pi^2}\lbrack\partial_{\mu, x}H_{\mathrm{mod}}, \partial_{\nu, x}H_{\mathrm{mod}}\rbrack.
\eea
Now we show that $\lbrack \partial_{\mu, x}H_{\mathrm{mod}}, \partial _{\nu, x} H_{\mathrm{mod}}\rbrack$ vanishes:
\bea
&&\lbrack\partial_{\mu, x} K, \partial_{\nu, x} K\rbrack
\nn\\
&=&
\frac{1}{2}\big(\partial_{\mu, x} \lbrack K, \partial_{\nu, x} K\rbrack 
-\lbrack K, \partial_{\mu, x}\partial_{\nu, x}K\rbrack
+\partial_{\nu, x}\lbrack \partial_{\mu, x} K, K\rbrack-\lbrack\partial_{\mu, x}\partial_{\nu, x}K, K\rbrack
\big)
\nn\\
&=&
\frac{1}{2}\big(\partial_{\mu, x} \lbrack K, \partial_{\nu, x} K\rbrack 
+\partial_{\nu, x}\lbrack \partial_{\mu, x} K, K\rbrack\big)
\nn\\
&=&\frac{1}{2}(-2\pi\partial_{\mu, x}\partial_{\nu, x}K+2\pi\partial_{\mu, x}\partial_{\nu, x}K)
\nn\\
&=&0.
\eea
The case for $-\lbrack V_{\mu, y}, V_{\nu, y}\rbrack$ is the same:
\bea
-\lbrack V_{\mu, y}, V_{\nu, y}\rbrack 
&\equiv& -\lbrack V_{\xd \xl_1}\at{\xl_1 = y^\mu}, V_{\xd \xl_2}\at{\xl_2 = y^\nu} \rbrack \nn\\
\nn\\
&=&
\frac{1}{4\pi^2}\lbrack\partial_{\mu, y}H_{\mathrm{mod}}, \partial_{\nu, y}H_{\mathrm{mod}}\rbrack.
\eea 
Hence this commutator vanishes as the $\lbrack V_{\mu, x}, V_{\nu, x}\rbrack$.
Finally, we calculate $-\lbrack V_{\mu, x}, V_{\nu, y}\rbrack$:
\bea
-\lbrack V_{\mu, x}, V_{\nu, y}\rbrack
&\equiv& -\lbrack V_{\xd \xl_1}\at{\xl_1 = x^\mu}, V_{\xd \xl_2}\at{\xl_2 = y^\nu} \rbrack \nn\\
&=&
-\frac{1}{4\pi^2}\lbrack\partial_{\mu, x}H_{\mathrm{mod}}, \partial_{\nu, y}H_{\mathrm{mod}}\rbrack.
\eea

\section{Berry Curvature and Riemann Curvature Tensor}
\label{sec:4}
\noindent
We first introduce the metric of a kinematic space, and then we construct the Berry curvature. The Berry curvature has the familiar property of the Riemann curvature, internal symmetry; skew symmetry; first Bianchi identity.
After we substitute the metric to the Berry curvature and Riemann curvature, they will give an equivalent result. Here we show an explicit calculation to the simplest example, CFT$_1$.

\subsection{Metric in the Kinematic Space}
\noindent
The spacetime interval on the kinematic space is \cite{deBoer:2016pqk}
\bea
ds^2=\frac{4}{(x-y)^2}\bigg(-\eta_{\mu\nu}+\frac{2(x_{\mu}-y_{\mu})(x_{\nu}-y_{\nu})}{(x-y)^2}\bigg)dx^{\mu}dy^{\nu},
\eea
where
\bea
\eta_{\mu\nu}\equiv\mathrm{diag}(-, +, +\cdots, +).
\eea
The vielbein is defined by
\bea
g_{\mu\nu}\equiv e_{\mu}{}^{a-}e_{\nu}{}^{b+}\eta_{ab}\eta_{+-},
\eea
where 
\bea
\eta_{+-}\equiv\frac{1}{2}, \qquad \eta_{++}=\eta_{--}=0;
\eea
\bea
\eta_{ab}\equiv\mathrm{diag}(-, +, +\cdots, +).
\eea
The internal indices are labeled by $a, b=0, 1, \cdots, d-1$.
\\

\noindent
Our choice of the vielbein is: 
\bea
e_{\mu}{}^{a-}\equiv\frac{2\sqrt{2}}{\sqrt{-(x-y)^2}}\delta_{\mu}{}^a; \qquad 
e_{\mu}{}^{a+}\equiv\frac{\sqrt{-(x-y)^2}}{\sqrt{2}}g_{\mu b}\eta^{ba}.
\eea
The inverse of the vielbeins are given by:
\bea
e_{a-}{}^{\mu}\equiv\frac{\sqrt{-(x-y)^2}}{2\sqrt{2}}\delta_a{}^{\mu}; \qquad
e_{a+}{}^{\mu}=\frac{\sqrt{2}}{\sqrt{-(x-y)^2}}\eta_{ab}g^{b\mu}.
\eea
\\

\noindent
We define: 
\bea
t\equiv\frac{1}{2}(x+y); \qquad z\equiv\frac{1}{2}(y-x)
\eea
when $d=1$. 
Then the spacetime interval becomes:
\bea
ds^2=-\frac{1}{z^2}\bigg(dt^2-dz^2-2(dt^2-dz^2)\bigg)=\frac{1}{z^2}(dt^2-dz^2).
\eea
This metric goes back to the dS$_2$ metric. 
The construction of the geometry in the kinematic space has an ambiguity on an overall sign \cite{deBoer:2016pqk}. 
Hence the overall sign in CFT$_1$ case is different from the higher-dimensional CFTs does not mean any inconsistency. 
Integration over a co-dimensional two surface in Lorentzian CFTs maps operators in real space to operators in a kinematic space \cite{Huang:2019wzc}. 
In Lorentzian CFT$_1$, the co-dimensional-two surface is a point. Hence the kinematic space of CFT$_1$ is AdS$_2$ \cite{Huang:2019wzc}. The metric in the kinematic space has a minus sign when $d=1$. 
\\

\noindent
Hence we will use the spacetime interval for $d>1$
\bea
ds^2=\frac{4}{(x-y)^2}\bigg(-\eta_{\mu\nu}+\frac{2(x_{\mu}-y_{\mu})(x_{\nu}-y_{\nu})}{(x-y)^2}\bigg)dx^{\mu}dy^{\nu}
\eea
and the spacetime interval for $d=1$
\bea
ds^2=-\frac{4}{(x-y)^2}\bigg(-\eta_{\mu\nu}+\frac{2(x_{\mu}-y_{\mu})(x_{\nu}-y_{\nu})}{(x-y)^2}\bigg)dx^{\mu}dy^{\nu}
\eea
to examine the equivalence between the Berry curvature and Riemann curvature tensor.

\subsection{Berry Curvature}
\noindent
We choose the Lie algebras of SO(2, $d$), $L_{+-}$, $L_{ab}$, and $L_{aj}$. The index $j$ is either $+$ or $-$. 
The number of non-trivial components is 1 for $L_{+-}$, $d(d-1)/2$ for $L_{ab}$, and $2d$ for $L_{aj}$. 
The sum of all numbers gives the degrees of freedom of SO(2, $d$): 
\bea
1+\frac{d(d-1)}{2}+2d=\frac{d^2+3d+2}{2}=\frac{(d+2)(d+1)}{2}=C^{d+2}_2.
\eea
The $L_{a\pm}$ is the eigenvector of the modular Hamiltonian with the eigenvalue $\pm 1$ respectively.
\\

\noindent
A matrix representation of the modular Hamiltonian is given by
\bea
H_{\mathrm{mod}}=-4\pi  L_{+-},
\eea
where 
\bea
(L_{+-})_{cd; jk}=-i(\eta_{+j}\eta_{-k}-\eta_{+k}\eta_{-j})\eta_{cd};
\eea
the derivative of the modular Hamiltonian is given by (according to their eigenvalues \er{algebradeform}):
\bea
\partial_{\mu, x}H_{\mathrm{mod}}\sim2\pi i e_{\mu}{}^{a-}L_{a-}; \qquad 
\partial_{\mu, y}H_{\mathrm{mod}}\sim2\pi i e_{\mu}{}^{a+}L_{a+}.\label{deriH}
\eea
\\

\noindent
To obtain the Berry curvature in the matrix representation, we use the commutator relation
\bea
\lbrack L_{aj}, L_{bk}\rbrack=i(\eta_{ab}L_{jk}+L_{ab}\eta_{jk}),
\eea
where 
\bea
(L_{ab})_{cd; jk}=-i(\eta_{ac}\eta_{bd}-\eta_{ad}\eta_{bc})\eta_{jk}.
\eea
We can calculate 
\bea
e_{\mu}^{a-}e_{\nu}^{b+}\lbrack L_{a-}, L_{b+}\rbrack
=i (2g_{\mu\nu} L_{-+}+e_{\mu}^{a-}e_{\nu}^{b+}L_{ab}\eta_{-+}).\label{eeLL}
\eea
to determine the coefficient ($\alpha$, see below) due to the
numerical factor \er{deriH} between the derivative of the modular Hamiltonian and the generator $L_{ai}$. More explicitly, it can be determined by comparing Eq. \er{eeLL}
and $\lbrack\partial_{\mu, y} H_{\mathrm{mod}}, \partial_{\nu, x} H_{\mathrm{mod}}\rbrack$. Because the metric appears, an overall sign for the metric will change the coefficient (note that
$\lbrack\partial_{\mu, y} H_{\mathrm{mod}}, \partial_{\nu, x} H_{\mathrm{mod}}\rbrack$ is independent
of the metric). 
The coefficient is $\alpha=1$ for CFT$_1$ and $\alpha=-1$ for the higher-dimensional CFTs. 
\\

\noindent
The Berry curvature is:
\bea
({\cal R}_{\mu+\nu-})_{\rho; j}{}^{\sigma; k}&\equiv&- (\lbrack V_{\mu, y}, V_{\nu, x}\rbrack)_{\rho; j}{}^{\sigma; k}
\nn\\
&=&-\frac{1}{4\pi^2}(\lbrack\partial_{\mu, y} H_{\mathrm{mod}}, \partial_{\nu, x} H_{\mathrm{mod}}\rbrack)_{\rho; j}{}^{\sigma; k}
\nn\\
&=&
-\alpha e_{\mu}{}^{a+}e_{\nu}{}^{b-}(\lbrack L_{a+}, L_{b-}\rbrack)_{\rho; j}{}^{\sigma; k}
\nn\\
&=&-i\alpha e_{\mu}{}^{a+}e_{\nu}{}^{b-}(\eta_{ab}L_{+-}+L_{ab}\eta_{+-})_{\rho; j}{}^{\sigma; k}.
\eea
Only when $j=k=\pm$, the Berry curvature does not vanish. The non-trivial components of the Berry curvature is written in terms the metric, given by:
\bea
({\cal R}_{\mu+\nu-})_{\rho; -}{}^{\sigma; -}
&=&-\alpha(g_{\mu\nu}\delta_{\rho}{}^{\sigma}+g_{\mu\rho}\delta_{\nu}{}^{\sigma}-g_{\mu b}\eta^{b\sigma}\eta_{\nu\rho});
\nn\\
({\cal R}_{\mu+\nu-})_{\rho; +}{}^{\sigma; +}
&=&-\alpha(-g_{\mu\nu}\delta_{\rho}{}^{\sigma}
+g_{\mu a}g_{\rho b}\eta^{ab}\eta_{\nu c}g^{c\sigma}
-\delta_{\mu}{}^{\sigma}g_{\nu\rho}).
\eea
We can also do a contraction to obtain the below equation:
\bea
({\cal R}_{\mu+\nu-})_{\rho; -;\sigma; +}
&=&
({\cal R}_{\mu+\nu-})_{\rho; -}{}^{\delta; -}g_{\delta\sigma}\eta_{-+}
\nn\\
&=&-\frac{\alpha}{2}(
g_{\mu\nu}g_{\rho\sigma}+g_{\mu\rho}g_{\nu\sigma}-g_{\mu a}\eta^{ab}g_{b\sigma}\eta_{\nu\rho});
\nn\\
({\cal R}_{\mu+\nu-})_{\rho; +;\sigma; -}
&=&
({\cal R}_{\mu+\nu-})_{\rho; +}{}^{\delta; +}g_{\delta\sigma}\eta_{+-}
\nn\\
&=&-\frac{\alpha}{2}(
-g_{\mu\nu}g_{\rho\sigma}+g_{\mu a}g_{\rho b}\eta^{ab}\eta_{\nu\sigma}-g_{\mu\sigma}g_{\nu\rho}).
\eea
It is easy to show the internal symmetry
\bea
({\cal R}_{\mu+\nu-})_{\rho; +;\sigma; -}=({\cal R}_{\rho+\sigma-})_{\mu; +;\nu; -},
\eea
the skew symmetry: 
\bea
({\cal R}_{\mu+\nu-})_{\rho;-;\sigma;+}=-({\cal R}_{\mu+\nu-})_{\sigma;+;\rho;-}
=-({\cal R}_{\nu-\mu+})_{\rho;-;\sigma;+},
\eea
and the first (algebraic) Bianchi identity
\bea
({\cal R}_{\mu+\nu-})_{\rho;+;\sigma;-}+({\cal R}_{\mu+\sigma-})_{\nu;-;\rho;+}=0.
\eea
Substituting the metric of the kinematic space into the Berry curvature and Riemann curvature tensor will show complete agreement in every dimension. 
We will give an explicit calculation to the most simple example, CFT$_1$ as a demonstration.

\subsection{Riemann Curvature Tensor}
\noindent
The Riemann curvature tensor is given by:
\bea
&&R^{\rho, i_{\rho}}{}_{\sigma, i_{\sigma}, \mu, i_{\mu}, \nu, i_{\nu}}
\nn\\
&\equiv&\partial_{\mu, i_{\mu}}\Gamma^{\rho, i_{\rho}}{}_{\nu, i_{\nu}, \sigma, i_{\sigma}}-\partial_{\nu, i_{\nu}}\Gamma^{\rho, i_{\rho}}{}_{\mu, i_{\mu}, \sigma, i_{\sigma}}
\nn\\
&&
+\Gamma^{\rho, i_{\rho}}{}_{\mu, i_{\mu}, \lambda, i_{\lambda}}\Gamma^{\lambda, i_{\lambda}}{}_{\nu, i_{\nu}, \sigma, i_{\sigma}}-\Gamma^{\rho, i_{\rho}}{}_{\nu, i_{\nu}, \lambda, i_{\lambda}}\Gamma^{\lambda, i_{\lambda}}{}_{\mu, i_{\mu}, \sigma, i_{\sigma}};
\nn\\
&&\Gamma^{\mu, i_{\mu}}{}_{\nu, i_{\nu}, \delta, i_{\delta}}
\nn\\
&\equiv&\frac{1}{2}g^{\mu, i_{\mu}, \lambda, i_{\lambda}}(\partial_{\delta, i_{\delta}}g_{\lambda, i_{\lambda}, \nu, i_{\nu}}+\partial_{\nu, i_{\nu}}g_{\lambda, i_{\lambda}, \delta, i_{\delta}}-\partial_{\lambda, i_{\lambda}}g_{\nu, i_{\nu}, \delta, i_{\delta}})
\eea
 in the kinematic space. The $\rho$ direction is for $x$ or $y$, labeled by the index $i_{\rho}$.
 For a comparison between the Riemann curvature tensor and Berry curvature, we directly calculate 
 \bea
 R_{\rho, i_{\rho}, \sigma, i_{\sigma}, \mu, i_{\mu}, \nu, i_{\nu}}\equiv \tilde{g}_{\rho, i_{\rho}, \delta, i_{\delta}}R^{\delta, i_{\delta}}{}_{\sigma, i_{\sigma}, \mu, i_{\mu}, \nu, i_{\nu}},
 \eea 
 where
 \bea
\tilde{g}_{\rho, i_{\rho}, \delta, i_{\delta}}\equiv g_{\rho, i_{\rho}, \delta, i_{\delta}}\eta_{i_{\rho} i_{\delta}},
 \eea
in which we do not sum over the indices, $i_{\rho}$ and $i_{\delta}$. 
The comparison will show an exact matching.

\subsection{CFT$_1$}
\noindent
We calculate the Berry curvature and Riemann curvature tensor in the kinematic space of CFT$_1$ and show their equivalence. 

\subsubsection{Berry Curvature}
\noindent
The non-vanishing component of the metric is 
\bea
g_{0+0-}=-\frac{2}{(x-y)(x-y)}.
\eea
We only have one non-trivial component:
\bea
({\cal R}_{0+0-})_{0+0-}&=&\frac{1}{2}(g_{0+0-}g_{0+0-}-g_{0+0-}g_{0+0-}\eta^{0-0-}\eta_{0-0-}+g_{0+0-}g_{0-0+})
\nn\\
&=&\frac{1}{2}g_{0+0-}g_{0-0+}
\nn\\
&=&\frac{2}{(x-y)(x-y)(x-y)(x-y)}.
\eea

\subsubsection{Riemann Curvature Tensor}
\noindent
We only have one non-trivial component:
\bea
R_{0+0-0+0-}&=&\eta_{+-}g_{0+0-}R^{0-}{}_{0-0+0-}=-\frac{1}{(x-y)(x-y)}\bigg(-\frac{2}{(x-y)(x-y)}\bigg)
\nn\\
&=&
\frac{2}{(x-y)(x-y)(x-y(x-y)},
\eea
in which we used:
\bea
&&
R^{0-}{}_{0-0+0-}
\nn\\
&=&\partial_{0+}\Gamma^{0-}{}_{0-0-}-\partial_{0-}\Gamma^{0-}{}_{0+0-}
\nn\\
&&
+\Gamma^{0-}{}_{0+0+}\Gamma^{0+}{}_{0-0-}+\Gamma^{0-}{}_{0+0-}\Gamma^{0-}{}_{0-0-}
\nn\\
&&
-\Gamma^{0-}{}_{0-0+}\Gamma^{0+}{}_{0+0-}-\Gamma^{0-}{}_{0-0-}\Gamma^{0-}{}_{0+0-}
\nn\\
&=&
\partial_{0+}\Gamma^{0-}{}_{0-0-}-\partial_{0-}\Gamma^{0-}{}_{0+0-}
+\Gamma^{0-}{}_{0+0+}\Gamma^{0+}{}_{0-0-}
-\Gamma^{0-}{}_{0-0+}\Gamma^{0+}{}_{0+0-}
\nn\\
&=&\partial_{0+}\Gamma^{0-}{}_{0-0-}
\nn\\
&=&-\frac{2}{(x-y)(x-y)},
\eea
where
\bea
\Gamma^{0-}{}_{0-0-}&=&
\frac{1}{2}g^{0-0+}(\partial_{0-}g_{0+0-}+\partial_{0-}g_{0+0-}-\partial_{0+}g_{0-0-})
=
g^{0-0+}\partial_{0-}g_{0+0-}
\nn\\
&=&
\frac{(x-y)(x-y)}{2}\bigg(-\frac{4}{(x-y)^3}\bigg)
=
-\frac{2}{x-y};
\nn
\eea
\bea
\Gamma^{0-}{}_{0+0-}&=&\frac{1}{2}g^{0-0+}(\partial_{0+}g_{0+0-}-\partial_{0+}g_{0+0-})=0;
\nn\\
\Gamma^{0-}{}_{0+0+}&=&0.
\eea
\\

\noindent
Hence Riemann curvature tensor matches the Berry curvature in the kinematic space of CFT$_1$.

\section{CFT$_1$ and Higher-Dimensional CFTs}
\label{sec:5}
\noindent
In this section, we compare CFT$_1$ to the higher-dimensional CFTs for the OPE block of a stress tensor. In CFT$_1$, the stress tensor is given by the Virasoro generator $L_{-2}$. We check the solution of the OPE block in Appendix \ref{appb}.

\subsection{OPE block}
\noindent
 The operator product expansion (OPE) of two operators ${\cal O}_j(x_1)$ and ${\cal O}_k(x_2)$ is given by
 \bea
 {\cal O}_j(x_1){\cal O}_k(x_2)=\sum_lC_{jkl}\big(x_1-x_2, \partial\big){\cal O}_l(x_2).
 \eea
 We then define the OPE block $B_l^{jk}(x_1, x_2)$ \cite{Czech:2016xec} as below
 \bea
  {\cal O}_j(x_1){\cal O}_k(x_2)\equiv |x_1-x_2|^{-\Delta_j-\Delta_k}\sum_l C_{jkl}\big(x_1-x_2, \partial\big)B_l^{jk}(x_1, x_2),
 \eea
 where 
 $\Delta_j$ and $\Delta_k$ are the scaling dimensions of the operators ${\cal O}_j$ and ${\cal O}_k$ respectively.

\subsection{Solution for CFT$_1$}
\noindent
 When we consider CFT$_1$, the OPE block satisfies \cite{Huang:2019wzc}
 \bea
 z^2(-\partial_z^2+\partial_t^2)B_k(\tau_1, \tau_2)=-\Delta_k(\Delta_k-1) B_k(\tau_1, \tau_2),
 \eea
 where 
 \bea
 t\equiv\frac{\tau_1+\tau_2}{2}, \qquad z\equiv\frac{\tau_1-\tau_2}{2}, \qquad \tau_1>\tau_2
 \eea
 or
 \bea
 t\equiv\frac{\tau_1+\tau_2}{2}, \qquad z\equiv\frac{\tau_2-\tau_1}{2}, \qquad \tau_2>\tau_1,
 \eea
 and $\Delta_k$ is the conformal dimension. A solution of the conformal block is
 \bea
 B_k(\tau_1, \tau_2)=\alpha_k\int_{\tau_1}^{\tau_2} dw\ \bigg(\frac{|w-\tau_2||w-\tau_1|}{|\tau_1-\tau_2|}\bigg)^{\Delta_k-1} {\cal O}_k(w), 
 \eea
 where $\alpha_k$ is an arbitrary constant for each $k$. The solution is checked in Appendix \ref{appb}.

\subsection{OPE Block of Stress Tensor}
\noindent
 The OPE block of a stress tensor in CFT$_1$ is given by
\bea
 B(\tau_1, \tau_2)\sim\int_{\tau_1}^{\tau_2} dw\ \frac{(\tau_2-w)(w-\tau_1)}{\tau_2-\tau_1} T(w), 
 \eea
 in which we assume $\tau_2>\tau_1$. 
 The conformal Killing vector in CFT$_1$ is given by 
 \bea
 K(w)=\frac{2\pi}{\tau_2-\tau_1}(\tau_2-w)(w-\tau_1).
 \eea
 Hence the OPE block of a stress tensor becomes
 \bea
 B(\tau_1, \tau_2)\sim\int_{\tau_1}^{\tau_2}dw\ K(w)T(w).
 \eea
 Therefore, we can find that the OPE block of a stress tensor in CFT$_1$ is similar to higher-dimensional CFTs except for the domain of the integration. 
 In the higher-dimensional CFTs, the integration of a modular Hamiltonian with a spherical entangling surface is over the ($d-1$)-dimensional spatial region. 
 The integration in CFT$_1$ is over a 1-dimensional time region. 
 Hence we do not expect that the OPE block of a stress tensor in CFT$_1$ can be related to a reduced density matrix \cite{Czech:2016xec, deBoer:2016pqk} as in the CFT$_2$. 
 However, the OPE block of a stress tensor in CFT$_1$ is still related to the AdS$_2$ Riemann curvature tensor \cite{Huang:2019wzc} because the operator dictionary already suggests that a stress tensor is dual to a bulk operator. 
 Hence the definition of a modular Hamiltonian should be useful for a generalization of holographic studies in CFTs.

\section{Discussion and Conclusion}
\label{sec:6}
\noindent
In this paper, we generalized the construction of the Berry curvature in the kinematic space \cite{Czech:2017zfq}, associated with a spherical entangling surface. 
We used the holonomy to derive the Berry curvature. 
The Berry curvature has the same properties as in the Riemann curvature: internal symmetry; skew symmetry; first Bianchi identity. 
The Berry curvature is also dual to the familiar Riemann curvature tensor in the kinematic space.
The procedure of the derivation is purely geometric. 
The algebra also gives a byproduct for the maximal modular scrambling modes, which relates quantum entanglement to quantum chaos.  
Finally, we discussed the difference between CFT$_1$ and the higher-dimensional CFTs from the OPE block \cite{Czech:2016xec, deBoer:2016pqk} of a stress tensor. 
\\

\noindent
The geometry of a kinematic space was constructed by symmetry. Therefore, an overall sign cannot be fixed. 
In other words, purely kinematic construction cannot determine the geometry. 
When we consider the CFT$_1$ case, the co-dimensional two surfaces in the bulk are a point. 
The integral geometry implies that the AdS$_2$ geometry should be the geometry of the kinematic space \cite{Huang:2019wzc}. 
The overall sign in the geometry affects the sign of the Berry curvature. 
The equivalence between the Berry curvature and Riemann curvature tensor can occur in both dS$_2$ and AdS$_2$ geometries. 
Hence determining the sign should include dynamical information like a bulk reconstruction of equations of motion. 
Reconstructing the bulk dynamics of a kinematic space is still a challenging direction. 
However, relating the Berry curvature to the Riemann geometry offers an alternative opinion to the kinematic space.
\\

\noindent
The algebra of the modular Hamiltonian with a spherical entangling surface shows the maximally chaotic modular scrambling modes. 
Because the derivation purely relies on the algebra, the saturation \cite{deBoer:2019uem} can be applied to any CFTs with a spherical entangling surface. 
This result implies that the only information of saturation cannot tell whether a theory is chaotic and holographic. 
However, chaotic information is not fully determined by the sensitivity of the initial condition. 
We still need to calculate other chaos quantities.
This direction should provide an exploration of a holographic study from the modular chaos \cite{deBoer:2019uem}.
\\

\noindent
We defined a modular Hamiltonian in CFT$_1$, which has a similar form to the higher-dimensional CFTs \cite{Huang:2019wzc}.
Because it only has time, one cannot use a division of space to define a reduced density matrix. 
Therefore, the modular Hamiltonian of a stress tensor in CFT$_1$ cannot be defined by a reduced density matrix. 
However, the variation of the modular Hamiltonian still shows the variation of the AdS$_2$ geometry \cite{Huang:2019wzc}. 
This possibly implies that the OPE block of a stress tensor is more fundamental than having an entanglement picture from a holographic picture.
Hence it is interesting to generalize a holographic study of the modular Hamiltonian. 

\section*{Acknowledgments}
\noindent
We would like to thank Bartlomiej Czech for his useful discussion.
\\

\noindent
Xing Huang was supported by the NSFC Grant No. 11947301 and the Double First-class University Construction Project of Northwest University. 
Chen-Te Ma was supported by the Post-Doctoral International Exchange Program; 
China Postdoctoral Science Foundation, Postdoctoral General Funding: Second Class (Grant No. 2019M652926); 
Science and Technology Program of Guangzhou (Grant No. 2019050001), and would like to thank Nan-Peng Ma for his encouragement.
\\

\noindent
We would like to thank the National Tsing Hua University and Yukawa Institute for Theoretical Physics at Kyoto University.
\\

\noindent
Discussions during the workshops, ``New Frontiers in String Theory'' and 
``NCTS Annual Theory Meeting 2019: Particles, Cosmology and Strings'', were useful to complete this work.

\appendix
\section{Derivation of Algebra of Modular Hamiltonian}
\label{appa}
\noindent
The first algebra 
\bea
\lbrack H_{\mathrm{mod}}, H_{\mathrm{mod}}\rbrack=0.
\eea
 is trivially satisfied.  
\\

\noindent
Now we calculate the second algebra $\lbrack H_{\mathrm{mod}}, \partial_{\nu, x}H_{\mathrm{mod}}\rbrack$. 
It is equivalent to calculating:
\bea
&&-K^{\rho}\partial_{\rho, w}\partial_{\nu, x} K^{\mu}+\partial_{\nu, x}K^{\rho}\partial_{\rho, w}K^{\mu}
\nn\\
&=&
-\frac{4\pi^2}{\big((y-x)^2\big)^2}
\big((y-w)^2(x^{\rho}-w^{\rho})-(x-w)^2(y^{\rho}-w^{\rho})\big)
\nn\\
&&
\times
\big(2(w_{\rho}-y_{\rho})\delta^{\mu}_{\nu}+2\eta_{\rho\nu}(y^{\mu}-w^{\mu})+2\delta^{\mu}_{\rho}(x_{\nu}-w_{\nu})\big)
\nn\\
&&
-2\frac{y_{\nu}-x_{\nu}}{(y-x)^2}K^{\rho}\partial_{\rho, w}K^{\mu}
\nn\\
&&
+\frac{4\pi^2}{\big((y-x)^2\big)^2}
\big((y-w)^2\delta^{\rho}_{\nu}-2(x_{\nu}-w_{\nu})(y^{\rho}-w^{\rho})\big)
\nn\\
&&
\times
\big(-(y-w)^2\delta^{\mu}_{\rho}+(x-w)^2\delta^{\mu}_{\rho}
-2(y_{\rho}-w_{\rho})(x^{\mu}-w^{\mu})+2(x_{\rho}-w_{\rho})(y^{\mu}-w^{\mu})\big)
\nn\\
&&
+2\frac{y_{\nu}-x_{\nu}}{(y-x)^2}K^{\rho}\partial_{\rho, w}K^{\mu}
\nn\\
&=&
-\frac{4\pi^2}{\big((y-x)^2\big)^2}
\big((y-w)^2(x^{\rho}-w^{\rho})-(x-w)^2(y^{\rho}-w^{\rho})\big)
\nn\\
&&
\times
\big(2(w_{\rho}-y_{\rho})\delta^{\mu}_{\nu}+2\eta_{\rho\nu}(y^{\mu}-w^{\mu})+2\delta^{\mu}_{\rho}(x_{\nu}-w_{\nu})\big)
\nn\\
&&
+\frac{4\pi^2}{\big((y-x)^2\big)^2}
\big((y-w)^2\delta^{\rho}_{\nu}-2(x_{\nu}-w_{\nu})(y^{\rho}-w^{\rho})\big)
\nn\\
&&
\times
\big(-(y-w)^2\delta^{\mu}_{\rho}+(x-w)^2\delta^{\mu}_{\rho}
-2(y_{\rho}-w_{\rho})(x^{\mu}-w^{\mu})+2(x_{\rho}-w_{\rho})(y^{\mu}-w^{\mu})\big)
\nn\\
&=&
-\frac{4\pi^2}{\big((y-x)^2\big)^2}
\big(
2(y-w)^2(x^{\rho}-w^{\rho})(w_{\rho}-y_{\rho})\delta^{\mu}_{\nu}
+2(y-w)^2(x_{\nu}-w_{\nu})(y^{\mu}-w^{\mu})
\nn\\
&&
+2(y-w)^2(x^{\mu}-w^{\mu})(x_{\nu}-w_{\nu})
\nn\\
&&
+2(x-w)^2(y-w)^2\delta^{\mu}_{\nu}
-2(x-w)^2(y_{\nu}-w_{\nu})(y^{\mu}-w^{\mu})
\nn\\
&&
-2(x-w)^2(y^{\mu}-w^{\mu})(x_{\nu}-w_{\nu})
\nn\\
&&
+\big((y-w)^2\big)^2\delta^{\mu}_{\nu}
-(y-w)^2(x-w)^2\delta^{\mu}_{\nu}
\nn\\
&&
+2(y-w)^2(y_{\nu}-w_{\nu})(x^{\mu}-w^{\mu})
-2(y-w)^2(x_{\nu}-w_{\nu})(y^{\mu}-w^{\mu})
\nn\\
&&
-2(y-w)^2(x_{\nu}-w_{\nu})(y^{\mu}-w^{\mu})
+2(x-w)^2(y^{\mu}-w^{\mu})(x_{\nu}-w_{\nu})
\nn\\
&&
-4(y-w)^2(x_{\nu}-w_{\nu})(x^{\mu}-w^{\mu})
+4(y^{\rho}-w^{\rho})(x_{\rho}-w_{\rho})(x_{\nu}-w_{\nu})(y^{\mu}-w^{\mu})
\nn\\
&=&
-\frac{4\pi^2}{\big((y-x)^2\big)}
\nn\\
&&
\times
\big((y-x)^2(y-w)^2\delta^{\mu}_{\nu}
-2\big((y-x)^2-(y-w)^2-(x-w)^2\big)(x_{\nu}-w_{\nu})(y^{\mu}-w^{\mu})
\nn\\
&&
-2(y-w)^2(x_{\nu}-w_{\nu})(x^{\mu}-w^{\mu})
-2(x-w)^2(y_{\nu}-w_{\nu})(y^{\rho}-w^{\mu})
\big)
\nn\\
&=&
2\pi\partial_{\nu, x}K^{\mu},
\eea
in which we used:
\bea
&&\partial_{\nu, x} K^{\rho}
\nn\\
&=&
-\frac{2\pi}{(y-x)^2}\big((y-x)^2\delta^{\rho}_{\nu}-2(x_{\nu}-w_{\nu})(y^{\rho}-w^{\rho})\big)
+\frac{2(y_{\nu}-x_{\nu})}{(y-x)^2}K^{\rho}
\nn\\
&=&
-\frac{2\pi}{(y-x)^2}\big((y-x)^2\delta^{\rho}_{\nu}-2(x_{\nu}-w_{\nu})(y^{\rho}-w^{\rho})\big)
\nn\\
&&
-\frac{4\pi(y_{\nu}-x_{\nu})}{\big((y-x)^2\big)^2}\big((y-w)^2(x^{\rho}-w^{\rho})-(x-w)^2(y^{\rho}-w^{\rho})\big)
\nn\\
&=&-\frac{2\pi}{\big((y-x)^2\big)^2}
\nn\\
&&
\times
\big((y-x)^2(y-w)^2\delta^{\rho}_{\nu}-2(y-x)^2(x_{\nu}-w_{\nu})(y^{\rho}-w^{\rho})
\nn\\
&&
+2(y-w)^2(y_{\nu}-x_{\nu})(x^{\rho}-w^{\rho})-2(x-w)^2(y_{\nu}-x_{\nu})(y^{\rho}-w^{\rho})\big)
\nn\\
&=&
\frac{2\pi}{\big((y-x)^2\big)^2}
\nn\\
&&
\times
\big(-(y-x)^2(y-w)^2\delta^{\rho}_{\nu}
+2(y-x)^2(x_{\nu}-w_{\nu})(y^{\rho}-w^{\rho})
\nn\\
&&
-2(y-w)^2(y_{\nu}-x_{\nu})(x^{\rho}-w^{\rho})
+2(x-w)^2(y_{\nu}-x_{\nu})(y^{\rho}-w^{\rho})\big)
\nn\\
&=&
\frac{2\pi}{\big((y-x)^2\big)^2}
\nn\\
&&
\times
\bigg(-(y-x)^2(y-w)^2\delta^{\rho}_{\nu}
+2\big((y-x)^2-(y-w)^2-(x-w)^2\big)(x_{\nu}-w_{\nu})(y^{\rho}-w^{\rho})
\nn\\
&&
+2(y-w)^2(x_{\nu}-w_{\nu})(x^{\rho}-w^{\rho})
+2(x-w)^2(y_{\nu}-w_{\nu})(y^{\rho}-w^{\rho})\bigg)
;
\nn\\
&&\partial_{\rho, w}K^{\mu}
\nn\\
&=&
-\frac{2\pi}{(y-x)^2}
\nn\\
&&\times
\big(-(y-w)^2\delta^{\mu}_{\rho}
+(x-w)^2\delta^{\mu}_{\rho}
\nn\\
&&
-2(y_{\rho}-w_{\rho})(x^{\mu}-w^{\mu})
+2(x_{\rho}-w_{\rho})(y^{\mu}-w^{\mu})\big).
\eea
Hence the second algebra is 
\bea
\lbrack H_{\mathrm{mod}}, \partial_{\nu, x} H_{\mathrm{mod}}\rbrack=-2\pi i\partial_{\nu, x}H_{\mathrm{mod}}.
\eea
\\

\noindent
The final algebra
\bea
\lbrack H_{\mathrm{mod}}, \partial_{\nu, y} H_{\mathrm{mod}}\rbrack=2\pi i\partial_{\nu, y}H_{\mathrm{mod}}.
\eea
 can be derived similarly.
 
 \section{Solution of OPE Block in CFT$_1$}
 \label{appb}
 \noindent
 The OPE block of CFT$_1$ satisfies \cite{Huang:2016bkp}
 \bea
 z^2(-\partial_z^2+\partial_t^2)B_k(\tau_1, \tau_2)=-\Delta_k(\Delta_k-1) B_k(\tau_1, \tau_2),
 \eea
 where 
 \bea
 t\equiv\frac{\tau_1+\tau_2}{2}, \qquad z\equiv\frac{\tau_2-\tau_1}{2}, \qquad \tau_2>\tau_1,
 \eea
 and $\Delta_k$ is conformal dimension. A solution of the OPE block is
 \bea
 B_k(\tau_1, \tau_2)=\alpha_k\int_{\tau_1}^{\tau_2} dw\ \bigg(\frac{|w-\tau_2||w-\tau_1|}{|\tau_1-\tau_2|}\bigg)^{\Delta_k-1} {\cal O}_k(w), 
 \eea
 where $\alpha_k$ is a constant for each $k$.
 \\ 
 
 \noindent
 Now we check the solution:
 \bea
 &&\bigg(\frac{|w-\tau_2||w-\tau_1|}{|\tau_1-\tau_2|}\bigg)^{\Delta_k-1}
 \nn\\
 &=&\bigg(\frac{(\tau_2-w)(w-\tau_1)}{\tau_2-\tau_1}\bigg)^{\Delta_k-1}
 \nn\\
&=&\bigg(\frac{-w^2+w(\tau_2+\tau_1)-\tau_1\tau_2}{\tau_2-\tau_1}\bigg)^{\Delta_k-1}
\nn\\
&=&\bigg(\frac{-w^2+w(\tau_2+\tau_1)-\tau_1\tau_2}{\tau_2-\tau_1}\bigg)^{\Delta_k-1}
\nn\\
&=&\bigg(\frac{-w^2+2wt+z^2-t^2}{2z}\bigg)^{\Delta_k-1}
\nn\\
&=&\bigg(\frac{z}{2}-\frac{t^2}{2z}+\frac{wt}{z}-\frac{w^2}{2z}\bigg)^{\Delta_k-1};
\nn
\eea
\bea
&&\partial_z\bigg\lbrack \bigg(\frac{|w-\tau_2||w-\tau_1|}{|\tau_1-\tau_2|}\bigg)^{\Delta_k-1}\bigg\rbrack
\nn\\
&=&(\Delta_k-1)\bigg(\frac{1}{2}+\frac{t^2}{2z^2}-\frac{wt}{z^2}+\frac{w^2}{2z^2}\bigg)\bigg(\frac{z}{2}-\frac{t^2}{2z}+\frac{wt}{z}-\frac{w^2}{2z}\bigg)^{\Delta_k-2};
\nn\\
&&\partial^2_z\bigg\lbrack \bigg(\frac{|w-\tau_2||w-\tau_1|}{|\tau_1-\tau_2|}\bigg)^{\Delta_k-1}\bigg\rbrack
\nn\\
&=&(\Delta_k-1)\bigg(-\frac{t^2}{z^3}+\frac{2wt}{z^3}-\frac{w^2}{z^3}\bigg)\bigg(\frac{z}{2}-\frac{t^2}{2z}+\frac{wt}{z}-\frac{w^2}{2z}\bigg)^{\Delta_k-2}
\nn\\
&&+(\Delta_k-1)(\Delta_k-2)\bigg(\frac{1}{2}+\frac{t^2}{2z^2}-\frac{wt}{z^2}+\frac{w^2}{2z^2}\bigg)^2\bigg(\frac{z}{2}-\frac{t^2}{2z}+\frac{wt}{z}-\frac{w^2}{2z}\bigg)^{\Delta_k-3};
\nn\\
&&\partial_t\bigg\lbrack \bigg(\frac{|w-\tau_2||w-\tau_1|}{|\tau_1-\tau_2|}\bigg)^{\Delta_k-1}\bigg\rbrack
\nn\\
&=&(\Delta_k-1)\bigg(-\frac{t}{z}+\frac{w}{z}\bigg)\bigg(\frac{z}{2}-\frac{t^2}{2z}+\frac{wt}{z}-\frac{w^2}{2z}\bigg)^{\Delta_k-2};
\nn\\
&&\partial^2_t\bigg\lbrack \bigg(\frac{|w-\tau_2||w-\tau_1|}{|\tau_1-\tau_2|}\bigg)^{\Delta_k-1}\bigg\rbrack
\nn\\
&=&-(\Delta_k-1)\frac{1}{z}\bigg(\frac{z}{2}-\frac{t^2}{2z}+\frac{wt}{z}-\frac{w^2}{2z}\bigg)^{\Delta_k-2}
\nn\\
&&+(\Delta_k-1)(\Delta_k-2)\bigg(-\frac{t}{z}+\frac{w}{z}\bigg)^2\bigg(\frac{z}{2}-\frac{t^2}{2z}+\frac{wt}{z}-\frac{w^2}{2z}\bigg)^{\Delta_k-3};
\nn
 \eea
 \bea
 &&z^2\big(-\partial_z^2+\partial_t^2\big)\bigg(\frac{|w-\tau_2||w-\tau_1|}{|\tau_1-\tau_2|}\bigg)^{\Delta_k-1}
 \nn\\
 &=&(\Delta_k-1)\bigg(\frac{t^2}{z}-\frac{2wt}{z}+\frac{w^2}{z}-z\bigg)\bigg(\frac{z}{2}-\frac{t^2}{2z}+\frac{wt}{z}-\frac{w^2}{2z}\bigg)^{\Delta_k-2}
 \nn\\
 &&+(\Delta_k-1)(\Delta_k-2)
 \nn\\
 &&\times\bigg(-\frac{z^2}{4}-\frac{t^4}{4z^2}-\frac{3w^2t^2}{2z^2}-\frac{w^4}{4z^2}+\frac{t^2}{2}-wt+\frac{w^2}{2}+\frac{wt^3}{z^2}+\frac{w^3t}{z^2}\bigg)
 \nn\\
 &&\times\bigg(\frac{z}{2}-\frac{t^2}{2z}+\frac{wt}{z}-\frac{w^2}{2z}\bigg)^{\Delta_k-3}
 \nn\\
 &=&-2(\Delta_k-1)\bigg(\frac{z}{2}-\frac{t^2}{2z}+\frac{wt}{z}-\frac{w^2}{2z}\bigg)^{\Delta_k-1}
 \nn\\
 &&-(\Delta_k-1)(\Delta_k-2)\bigg(\frac{z}{2}-\frac{t^2}{2z}+\frac{wt}{z}-\frac{w^2}{2z}\bigg)^{\Delta_k-1}
 \nn\\
 &=&-\Delta_k(\Delta_k-1)\bigg(\frac{z}{2}-\frac{t^2}{2z}+\frac{wt}{z}-\frac{w^2}{2z}\bigg)^{\Delta_k-1}.
 \eea
 Hence we conclude that the solution satisfies the below equation
  \bea
 z^2(-\partial_z^2+\partial_t^2)B_k(\tau_1, \tau_2)=-\Delta_k(\Delta_k-1) B_k(\tau_1, \tau_2).
 \eea


\baselineskip 22pt
\begin{thebibliography}{99}

\bibitem{tHooft:1996rdg} 
  G.~'t Hooft,
  ``The Scattering matrix approach for the quantum black hole: An Overview,''
  Int.\ J.\ Mod.\ Phys.\ A {\bf 11}, 4623 (1996)
  doi:10.1142/S0217751X96002145
  [gr-qc/9607022].

\bibitem{Strominger:2017zoo} 
  A.~Strominger,
  ``Lectures on the Infrared Structure of Gravity and Gauge Theory,''
  arXiv:1703.05448 [hep-th].

\bibitem{Maldacena:1997re} 
  J.~M.~Maldacena,
  ``The Large N limit of superconformal field theories and supergravity,''
  Int.\ J.\ Theor.\ Phys.\  {\bf 38}, 1113 (1999)
  [Adv.\ Theor.\ Math.\ Phys.\  {\bf 2}, 231 (1998)]
  doi:10.1023/A:1026654312961, 10.4310/ATMP.1998.v2.n2.a1
  [hep-th/9711200].
  
\bibitem{Ryu:2006bv} 
  S.~Ryu and T.~Takayanagi,
  ``Holographic derivation of entanglement entropy from AdS/CFT,''
  Phys.\ Rev.\ Lett.\  {\bf 96}, 181602 (2006)
  doi:10.1103/PhysRevLett.96.181602
  [hep-th/0603001].
  
\bibitem{Casini:2011kv} 
  H.~Casini, M.~Huerta and R.~C.~Myers,
  ``Towards a derivation of holographic entanglement entropy,''
  JHEP {\bf 1105}, 036 (2011)
  doi:10.1007/JHEP05(2011)036
  [arXiv:1102.0440 [hep-th]].
  
\bibitem{Ma:2015xes} 
  C.~T.~Ma,
  ``Entanglement with Centers,''
  JHEP {\bf 1601}, 070 (2016)
  doi:10.1007/JHEP01(2016)070
  [arXiv:1511.02671 [hep-th]].
  
\bibitem{Huang:2016bkp}
X.~Huang and C.~T.~Ma,
``Analysis of the Entanglement with Centers,''
J. Stat. Mech. \textbf{2005}, 053101 (2020)
doi:10.1088/1742-5468/ab7c63
[arXiv:1607.06750 [hep-th]].
  
\bibitem{Ma:2016deg} 
  C.~T.~Ma,
  ``Discussion of Entanglement Entropy in Quantum Gravity,''
  Fortsch.\ Phys.\  {\bf 66}, no. 2, 1700095 (2018)
  doi:10.1002/prop.201700095
  [arXiv:1609.03651 [hep-th]].
  
\bibitem{Ma:2016pah} 
  C.~T.~Ma,
  ``Theoretical Properties of Entropy in a Strong Coupling Region,''
  Class.\ Quant.\ Grav.\  {\bf 35}, no. 23, 235011 (2018)
  doi:10.1088/1361-6382/aaec3b
  [arXiv:1609.04550 [hep-th]].
  
\bibitem{Ma:2018efs} 
  C.~T.~Ma,
  ``Parity Anomaly and Duality Web,''
  Fortsch.\ Phys.\  {\bf 66}, no. 8-9, 1800045 (2018)
  doi:10.1002/prop.201800045
  [arXiv:1802.08959 [hep-th]].
  
\bibitem{Czech:2016xec} 
  B.~Czech, L.~Lamprou, S.~McCandlish, B.~Mosk and J.~Sully,
  ``A Stereoscopic Look into the Bulk,''
  JHEP {\bf 1607}, 129 (2016)
  doi:10.1007/JHEP07(2016)129
  [arXiv:1604.03110 [hep-th]].

\bibitem{deBoer:2016pqk} 
  J.~de Boer, F.~M.~Haehl, M.~P.~Heller and R.~C.~Myers,
  ``Entanglement, holography and causal diamonds,''
  JHEP {\bf 1608}, 162 (2016)
  doi:10.1007/JHEP08(2016)162
  [arXiv:1606.03307 [hep-th]].
  
\bibitem{Witten:2018lha} 
  E.~Witten,
  ``APS Medal for Exceptional Achievement in Research: Invited article on entanglement properties of quantum field theory,''
  Rev.\ Mod.\ Phys.\  {\bf 90}, no. 4, 045003 (2018)
  doi:10.1103/RevModPhys.90.045003
  [arXiv:1803.04993 [hep-th]].
  
    
\bibitem{Dolan:2003hv} 
  F.~A.~Dolan and H.~Osborn,
  ``Conformal partial waves and the operator product expansion,''
  Nucl.\ Phys.\ B {\bf 678}, 491 (2004)
  doi:10.1016/j.nuclphysb.2003.11.016
  [hep-th/0309180].
  
\bibitem{SimmonsDuffin:2012uy} 
  D.~Simmons-Duffin,
  ``Projectors, Shadows, and Conformal Blocks,''
  JHEP {\bf 1404}, 146 (2014)
  doi:10.1007/JHEP04(2014)146
  [arXiv:1204.3894 [hep-th]].
  
\bibitem{Fitzpatrick:2015foa} 
  A.~L.~Fitzpatrick, J.~Kaplan, M.~T.~Walters and J.~Wang,
  ``Hawking from Catalan,''
  JHEP {\bf 1605}, 069 (2016)
  doi:10.1007/JHEP05(2016)069
  [arXiv:1510.00014 [hep-th]].
  
\bibitem{daCunha:2016crm} 
  B.~Carneiro da Cunha and M.~Guica,
  ``Exploring the BTZ bulk with boundary conformal blocks,''
  arXiv:1604.07383 [hep-th].

\bibitem{Guica:2016pid} 
  M.~Guica,
  ``Bulk fields from the boundary OPE,''
  arXiv:1610.08952 [hep-th].

\bibitem{Fitzpatrick:2016mtp} 
  A.~L.~Fitzpatrick, J.~Kaplan, D.~Li and J.~Wang,
  ``Exact Virasoro Blocks from Wilson Lines and Background-Independent Operators,''
  JHEP {\bf 1707}, 092 (2017)
  doi:10.1007/JHEP07(2017)092
  [arXiv:1612.06385 [hep-th]].
  
\bibitem{Hijano:2015zsa} 
  E.~Hijano, P.~Kraus, E.~Perlmutter and R.~Snively,
  ``Witten Diagrams Revisited: The AdS Geometry of Conformal Blocks,''
  JHEP {\bf 1601}, 146 (2016)
  doi:10.1007/JHEP01(2016)146
  [arXiv:1508.00501 [hep-th]].
  
\bibitem{Czech:2017zfq} 
  B.~Czech, L.~Lamprou, S.~Mccandlish and J.~Sully,
  ``Modular Berry Connection for Entangled Subregions in AdS/CFT,''
  Phys.\ Rev.\ Lett.\  {\bf 120}, no. 9, 091601 (2018)
  doi:10.1103/PhysRevLett.120.091601
  [arXiv:1712.07123 [hep-th]].

\bibitem{Czech:2019vih} 
  B.~Czech, J.~De Boer, D.~Ge and L.~Lamprou,
  ``A modular sewing kit for entanglement wedges,''
  JHEP {\bf 1911}, 094 (2019)
  doi:10.1007/JHEP11(2019)094
  [arXiv:1903.04493 [hep-th]].
  
\bibitem{Huang:2019wzc} 
  X.~Huang and C.~T.~Ma,
  ``The Probe of Curvature in the Lorentzian AdS$_2$/CFT$_1$ Correspondence,''
  Phys.\ Lett.\ B {\bf 798}, 134936 (2019)
  doi:10.1016/j.physletb.2019.134936
  [arXiv:1907.01422 [hep-th]].
  
\bibitem{Maldacena:2016hyu} 
  J.~Maldacena and D.~Stanford,
  ``Remarks on the Sachdev-Ye-Kitaev model,''
  Phys.\ Rev.\ D {\bf 94}, no. 10, 106002 (2016)
  doi:10.1103/PhysRevD.94.106002
  [arXiv:1604.07818 [hep-th]].
  
\bibitem{Callebaut:2018xfu} 
  N.~Callebaut,
  ``The gravitational dynamics of kinematic space,''
  JHEP {\bf 1902}, 153 (2019)
  doi:10.1007/JHEP02(2019)153
  [arXiv:1808.10431 [hep-th]].
  
\bibitem{Blommaert:2019hjr} 
  A.~Blommaert, T.~G.~Mertens and H.~Verschelde,
  ``Clocks and Rods in Jackiw-Teitelboim Quantum Gravity,''
  JHEP {\bf 1909}, 060 (2019)
  doi:10.1007/JHEP09(2019)060
  [arXiv:1902.11194 [hep-th]].
    
\bibitem{Shenker:2013pqa} 
  S.~H.~Shenker and D.~Stanford,
  ``Black holes and the butterfly effect,''
  JHEP {\bf 1403}, 067 (2014)
  doi:10.1007/JHEP03(2014)067
  [arXiv:1306.0622 [hep-th]].
  
\bibitem{Maldacena:2015waa} 
  J.~Maldacena, S.~H.~Shenker and D.~Stanford,
  ``A bound on chaos,''
  JHEP {\bf 1608}, 106 (2016)
  doi:10.1007/JHEP08(2016)106
  [arXiv:1503.01409 [hep-th]].
  
\bibitem{Perlmutter:2016pkf} 
  E.~Perlmutter,
  ``Bounding the Space of Holographic CFTs with Chaos,''
  JHEP {\bf 1610}, 069 (2016)
  doi:10.1007/JHEP10(2016)069
  [arXiv:1602.08272 [hep-th]].
  
\bibitem{Jensen:2016pah} 
  K.~Jensen,
  ``Chaos in AdS$_2$ Holography,''
  Phys.\ Rev.\ Lett.\  {\bf 117}, no. 11, 111601 (2016)
  doi:10.1103/PhysRevLett.117.111601
  [arXiv:1605.06098 [hep-th]].

\bibitem{Maldacena:2016upp} 
  J.~Maldacena, D.~Stanford and Z.~Yang,
  ``Conformal symmetry and its breaking in two dimensional Nearly Anti-de-Sitter space,''
  PTEP {\bf 2016}, no. 12, 12C104 (2016)
  doi:10.1093/ptep/ptw124
  [arXiv:1606.01857 [hep-th]].
  
\bibitem{deBoer:2019uem}
J.~De Boer and L.~Lamprou,
``Holographic Order from Modular Chaos,''
JHEP \textbf{06}, 024 (2020)
doi:10.1007/JHEP06(2020)024
[arXiv:1912.02810 [hep-th]].
        
\bibitem{Casini:2017roe} 
  H.~Casini, E.~Teste and G.~Torroba,
  ``Modular Hamiltonians on the null plane and the Markov property of the vacuum state,''
  J.\ Phys.\ A {\bf 50}, no. 36, 364001 (2017)
  doi:10.1088/1751-8121/aa7eaa
  [arXiv:1703.10656 [hep-th]].
                                        
\end{thebibliography}
\end{document}